# The Impact of China's Economic Growth on Poverty Alleviation: From Absolute to Relative Poverty

Yixun Kang[1,*] and Ying Li[2]
[1]Data Science Institute, Brown University, RI, USA
[2]Division of the Social Science, The University of Chicago, IL, USA
[*]Corresponding Author: e-mail: yixun_kang@brown.edu

**Abstract**

This paper investigates the extent to which China's economic growth and development influence poverty levels, focusing on the dichotomy between absolute and relative poverty. Leveraging data from sources like the World Bank, Statista, and Macrotrends, and employing economic frameworks such as the Lewis Model, Poverty Headcount Ratio, and Gini Coefficient, the study examines China's transformation from combating absolute poverty to addressing relative poverty. The findings highlight that robust economic growth from 2011 to 2022, driven by urban development and rural infrastructure investments, successfully eradicated absolute poverty and elevated rural incomes. However, this progress also exacerbated income inequality, as evidenced by a rising Gini Coefficient, complicating efforts to alleviate relative poverty. Through multidimensional analyses encompassing regional disparities, migration patterns, educational access, and societal factors, the paper underscores the dual impact of economic development on poverty alleviation. It concludes by advocating for policies that balance economic growth with equitable resource distribution to tackle persistent relative poverty and foster sustainable development.

## 1 Introduction

China is the world's most populous country and one of the largest economies. China has experienced rapid economic growth in recent decades, lifting millions of people out of poverty and emerging as a global economic powerhouse. By the end of 2020, 98.99 million rural populations had completely lifted themselves out of absolute poverty, aligning with the prevailing poverty standards [1]. Eliminating absolute poverty in rural China marks a significant transition in poverty alleviation efforts, shifting the focus from absolute poverty to relative poverty.

The essence of relative poverty is inequality, and its fundamental cause stems from inequitable income distribution and resource allocation. The rapid development of the urban economy has resulted in a substantial disparity between the per capita income of urban residents and the absolute poverty line in rural areas. The revenue generated from income tax paid by urban residents could be allocated toward developing rural infrastructure. Although urban economic growth has contributed to a decline in the absolute poverty rate by raising the income of urban residents, it has not effectively alleviated the relative poverty rate [2].

Taking into account China's unique national circumstances, this paper uses various economic concepts, such as the Lewis Model, Poverty Headcount Ratio, Poverty Gap Index, and Gini Coefficient, and statistics obtained from the World Bank, Statista and Macrotrends to address the following research question: To what extent does China's economic growth and development impact poverty levels?

The subsequent parts of this research thesis are organized as follows: Part 2 is the literature review; Part 3 introduces data sources and all the economic concepts used in the study; Part 4 contains analyses of absolute poverty and relative poverty; Part 5 is the conclusion.

## 2 Literature Review

China is currently undergoing a transition from addressing absolute poverty to alleviating relative poverty. As relative poverty is rooted in social inequality, it is challenging for market mechanisms alone to tackle this issue effectively. Additionally, the complex nature of urban relative poverty in China, coupled with the absence of a standardized urban relative poverty indicator, could impede government intervention [2].

Therefore, as scholars' understanding of urban poverty deepens, they have begun to integrate the concept of relative poverty with urban poverty: Yang Yang and Ma Xiao have discovered that migrant populations experience higher levels of poverty compared to urban residents; Yang Ge argues that the growing number of migrants has exacerbated the level of relative poverty in cities; Shen Yangyang proposes the implementation of relative poverty thresholds in both urban and rural areas, where the threshold for relative poverty is defined as 40% of the median income among residents [3][4][5]. Meanwhile, Chinese scholars have progressively shifted from a unidimensional to a multidimensional approach, using the Alkire-Foster method, when it comes to assessing urban poverty: Zhang Quanhong and Zhou Qiang conducted a comparative assessment of urban and rural poverty, considering five dimensions: education, health, employment, living conditions, and public services [6].

The impact of China's economic growth on relative and absolute poverty can vary depending on different concepts of poverty. Therefore, this thesis separates absolute poverty from relative poverty in addressing the research question and uses a multidimensional approach in analyzing the impact of economic growth on relative poverty. In the multidimensional relative poverty analysis, this study takes regional, migration, educational and personal factors into account.

## 3 Methodology

### 3.1 Data Source

This paper uses data provided by several reputable sources: the World Bank, Statista and Macrotrends, as well as official statistics derived from the Outline of China's Rural Poverty Alleviation and Development. In order to better concentrate on the impact of economic development on poverty alleviation, statistics from 2011 and 2022 are selected. Also, data from 1990 and 2019 are used as a reference for economic growth and poverty reduction, and income distribution, respectively.

### 3.2 Economics Concepts

There are multiple approaches to measure poverty, arising from diverse interpretations of the concept itself as well as differences in the poverty axioms. In this paper, $y = (y_1, y_2, \ldots, y_n)$, where $n$ is the sample size, and $y_i$ is the income level of individual $i$. The poverty threshold $z$ is defined by function $z: \mathbb{R} \to \mathbb{R}$. $\bar{y} = \sum_{i=1}^{n} \frac{y_i}{n}$ is the average income of the population. If individual $i$ has income $y_i < z(\bar{y})$, then individual $i$ is defined as impoverished. The amount of impoverished individuals is defined by $q(y)$.

### 3.2.1 Lewis Model

The Lewis Model, also known as the Dual-sector Model, was first proposed by Sir William Arthur Lewis in 1954. It is a theory of economic development in which surplus labor from the traditional agriculture sector is transferred to the modern industrial sector, whose expansion over time absorbs the extra labor and fosters industrialization and long-term development.

In the model, the traditional agriculture sector is characterized by low wages, labor surplus, and low productivity. In contrast, the modern industrial sector is characterized by higher wage rates, higher marginal productivity, and higher initial quantity demanded for labor. Moreover, in the modern industrial sector, investment and capital formation are possible over time.

### 3.2.2 Absolute Poverty and Relative Poverty

The concept of absolute poverty originated from Rowntree's theory of minimum budget level for subsistence. This approach measures poverty using a "basket of goods" that meets the minimum requirements for sustaining physical well-being [7]. Until 1979, poverty was commonly assessed by establishing a minimum economic threshold required to meet essential survival necessities, such as food, housing, and transportation. Subsequently, Townsend introduced the concept of relative poverty, which refers to a social phenomenon where individuals or groups experience relative deprivation in terms of resources, opportunities, and rights compared to a specific reference group. This deprivation leads to limited opportunities, restricted choice, diminished possibilities for income generation, and a lower quality of life [8].

### 3.2.3 Poverty Headcount Ratio (HCR)

The Poverty Headcount Ratio (HCR) is the earliest and a widely used poverty indicator. Given a certain poverty threshold $z$, HCR measures the proportion of the population $q(y)$ living below that line relative to the total population [9]. It is used to assess the extent of poverty. However, this indicator violates the poverty axioms of monotonicity and transfer sensitivity and lacks sensitivity toward the income distribution among the impoverished population [10].

$$HCR = \frac{q(y)}{n}$$

### 3.2.4 Poverty Gap Index (PGI)

The Poverty Gap Index (PGI) provides a thorough evaluation of the degree of deprivation that the impoverished population faces. By calculating the income shortfall of individuals below the poverty line, this indicator measures the depth of poverty, partially compensating for HCR's limitations in capturing poverty severity.

$$PGI = \frac{1}{n} \sum_{i=1}^{q(y)} \left(\frac{z - y_i}{z}\right)$$

### 3.2.5 Lorenz Curve ($L(p)$) and Gini Coefficient

The Lorenz Curve is a graphical depiction of income or wealth distribution. It displays the proportion $L(p)$ of overall income or wealth held by the bottom $p$ percentile of individuals. The x-axis represents the cumulative percentage of individuals, while the y-axis represents the cumulative percentage of income. A 45% line is also constructed with the Lorenz Curve, representing the tax system with no taxes imposed.

The Gini Coefficient is a statistical measure of distribution used to depict income, wealth, or consumption inequality within a country or a social group [11]. It is the most famous inequality indicator. A Gini Coefficient of 0 reflects perfect equality, where all incomes are equally distributed among individuals. A Gini Coefficient of 1 reflects complete inequality, where the top individual has all the income.

### 3.2.6 Education Index and Gross Enrollment Rate (GER)

The Education Index measures educational attainment, using the formula $EI = \frac{\frac{EYS}{18} + \frac{MYS}{15}}{2}$, where $EI$ is the Education Index, $EYS$ is the expected years of schooling, and $MYS$ is the mean years of schooling.

The Gross Enrollment Rate (GER) is the total enrollment in a specific level of education, regardless of age, expressed as a percentage of the population in the official age group corresponding to this level of education. The levels of education contain, primary, secondary, and tertiary.

## 4 Analysis

### 4.1 Absolute Poverty Analysis

At the beginning of 2011, the Chinese government outlined two fundamental strategies for poverty alleviation. Firstly, the government aims to inspire and assist individuals capable of lifting themselves from poverty through their endeavors [12]. Secondly, they are keen on strengthening the self-development skills of the impoverished to enhance their income-generating capabilities [12]. These two principles emphasize not just the increase of individual income but also the boost of sustainable, income-generating capacities. Throughout the Chinese government's efforts on developing infrastructure in rural areas, these two guiding principles were achieved by the end of 2018 [12].

### 4.1.1 Income Increment

In 2020, China accomplished a moderately prosperous society in all respects. From 2011 to 2020, China's GDP increased from $7.55 trillion in 2011 to $14.69 trillion in 2020 [13]. During this decade, China's economy developed rapidly, primarily driven by the Internet. This progress resulted in notable advancements in sectors like e-commerce, live streaming, and short videos, significantly impacting the real economy. Rural individual businesses have the opportunity to leverage the Internet for product sales, eliminating the necessity of entering physical markets. This online sales approach enables direct transactions, particularly for agricultural crops, effectively reducing freight costs and the risk of spoilage. Consequently, e-commerce targets a broader customer base, increasing sales and income.

Meanwhile, China was making concerted efforts to develop coastal cities and major urban centers like Beijing, Shanghai, Guangzhou, and Shenzhen. This urban development has provided abundant employment opportunities, attracting rural residents to migrate to cities. Urban development has contributed to an increase in the per capita income of urban residents. In 2011 and 2020, the average annual per capita income of urban residents reached $3316 (￥21427) and $6353 (￥43834), respectively [14]. Moreover, in 2020, urban residents' average daily per capita income was approximately 7.5 times higher than the national poverty threshold of $2.3 per day [15]. Increased income leads to higher tax revenue. Therefore, more money can be used to alleviate poverty. In 2020, the Chinese government invested $20.5 billion (￥141.6 billion) in poverty alleviation, developing the infrastructure in impoverished rural areas [16]. This includes access to clean drinking water, electricity, and transportation, as well as enhancing basic healthcare, education, and social security services in these regions. The development of water and electrical facilities and transportation created more job opportunities for rural residents, thereby shrinking the surplus labor in rural areas, enhancing marginal productivity, and further extending income levels.

### 4.1.2 Income-Generating Capabilities Enhancement

China's investment in infrastructure from 2011 to 2020 has had a profound impact on its Education Index, which rose from 0.63 to 0.86 [17]. This rise implies an enhanced return on education. Improved access to compulsory education enables children from rural areas to potentially leave their communities, pursue higher education in cities, and ultimately secure high-paying jobs. Simultaneously, the establishment of technical schools in rural areas aims to uplift the skill set of rural residents. This means that an individual, previously confined to basic agricultural work due to a lack of skills, can acquire engineering-related knowledge and subsequently qualify for better-paying positions.

The improvement in rural basic education has alleviated the situation where rural students have no school to attend or cannot afford schooling. Meanwhile, in labor-surplus rural areas, individuals, after receiving technical guidance, can separate themselves from the concentrated agricultural sector and engage in less labor-intensive industries or move to cities for employment. Therefore, the improvement raises the income in rural areas and enables individuals to help their families escape poverty.

### 4.1.3 Discussion

From 2011 to 2020, the per capita income of rural residents rose from $1144 (￥7394) to $2483 (￥17131) [14]. China stimulated development in rural areas by conducting aggressive urban growth and investing in basic rural infrastructure. This investment improved education levels among the impoverished rural population, subsequently opening opportunities for high-paying jobs. Additionally, the Chinese government provided substantial subsidies to severely impoverished rural residents.

A line chart containing GDP, HCR, and PGI for 1990, 2011, and 2020 was constructed to visualize their relationship. It should be noted that the poverty threshold in 1990 stood at $1.9 per day, while in 2011 and 2020, it increased to $2.3 per day. Therefore, both HCR and PGI in 1990 should be higher if using the current poverty threshold.

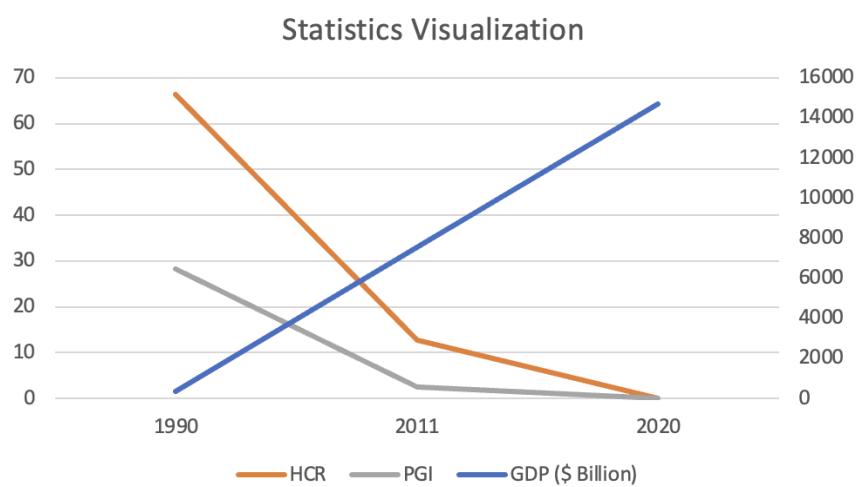

Figure 1: Statistics Visualization
Source: Created by Author

As illustrated in Figure 1, it is evident that there exists a negative correlation between GDP and both HCR and PGI. Between 1990 and 2011, both HCR and PGI declined significantly.

This trend of decline persisted into the next decade, although the pace of reduction slowed. Over this decade, China experienced a fall in HCR, from 12.7 to 0.1, and PGI, from 2.5 to 0.1, resulting in the complete elimination of absolute poverty by 2020 [18][19]. Rising GDP also significantly contributed to the reduction of relative poverty, from 58.9 to 13 [20][15]. Notably, GDP growth served as the primary factor in alleviating absolute poverty.

4.2 Relative Poverty Analysis

In order to further decrease China's relative poverty to a new minimum, China may need to adjust its economic strategies. Instead of exclusively focusing on GDP growth, a transition towards policies that promote wealth redistribution could be more effective. From 2011 to 2020, China's Gini Coefficient has risen from 42.4 to 57.1, indicating increased income inequality [21][22]. This suggests that while GDP growth has helped alleviate absolute poverty, it may not be as effective at tackling relative poverty due to the disparities in resource allocation.

4.2.1 Regional Factor

Due to the geographical diversity of China, the economies vary greatly from region to region, adding complexity to the effective redistribution of social wealth. Coastal cities in China have experienced remarkable economic growth compared to their inland counterparts, primarily attributed to their well-connected transportation for fluent international trade. Simultaneously, the urbanization of these coastal cities has presented a wide range of opportunities, benefiting both the impoverished and the affluent segments of society. However, individuals with greater financial privilege hold a more advantageous position when capitalizing on these opportunities. Their higher education levels and greater assets enable them to effectively leverage their existing businesses, resulting in dramatically increased income levels and exacerbated income gaps.

Before COVID-19, rural regions' total disposable income reached $1.2 trillion (￥8.8 trillion), while in urban areas, total disposable income achieved $5.21 trillion (￥36 trillion) [14]. Although urban regions had a lower Gini Coefficient of 0.35, the substantial disparity between rural total income and urban total income resulted in a more in-depth urban income gap between low percentiles and high percentiles [23]. Although economic growth has effectively reduced relative poverty, the inequality in income distribution has intensified due to the same economic expansion, thereby resulting in further relative poverty.

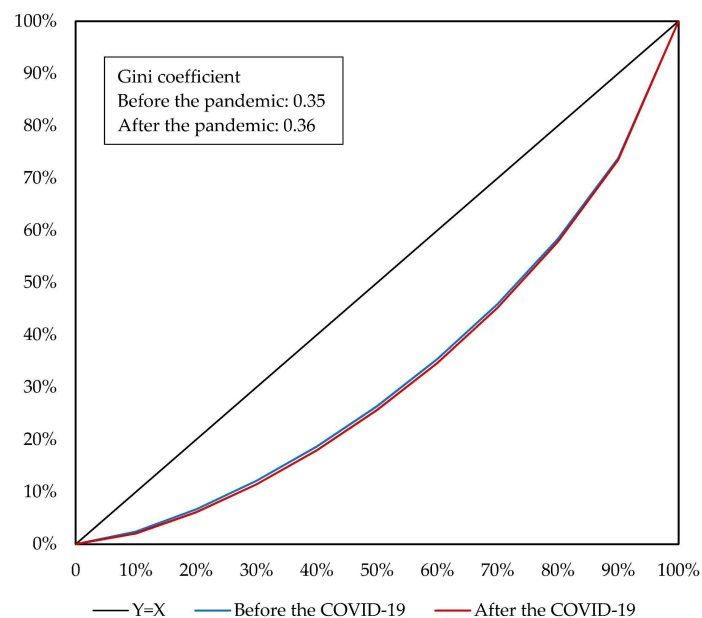

Figure 2: Gini Coefficient and Lorenz Curve of Urban Residents
Source: Zhang Qi, The Unequal Effect of the COVID-19 Pandemic on the Labour Market and Income Inequality in China [23]

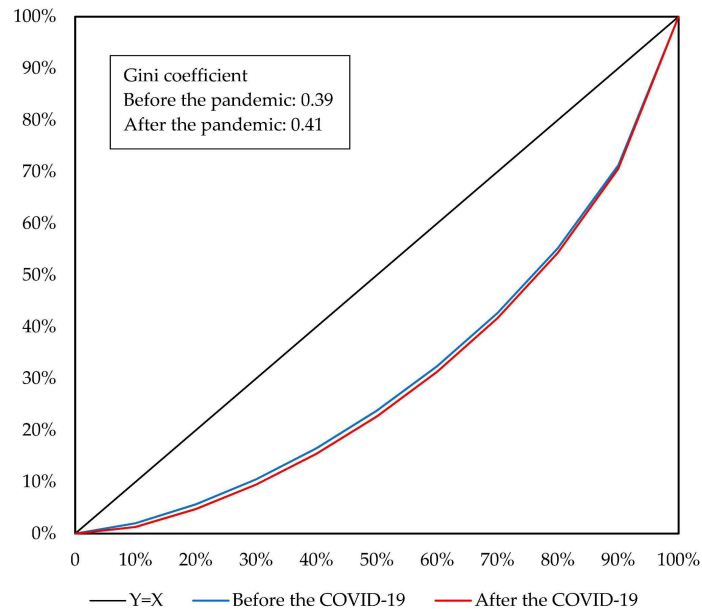

Figure 3: Gini Coefficient and Lorenz Curve of Rural Residents
Source: Zhang Qi, The Unequal Effect of the COVID-19 Pandemic on the Labour Market and Income Inequality in China [23]

Aside from regional income inequality, factors contributing to poverty also differ from region to region. In middle- and high-income areas, including major economic centers and the capital region, relative poverty may arise from business failures or insufficient family labor. In addition to these reasons, in low-income regions, such as remote mountain areas, relative poverty is exacerbated by regional economic underdevelopment and limited resources. China's national economic growth fails to effectively address these regional variations. The strategy of using urban development as a stimulus for rural economies exhibits certain drawbacks as it overlooks the unique contributing factors to poverty in rural areas. For instance, while rural infrastructure development has managed to satisfy basic needs like electricity, clean drinking water, and transportation, other pressing issues such as those associated with agricultural productivity continue to require further attention.

### 4.2.2 Migration Factor

The urbanization of China's major cities has highlighted evident disparities in the household (hukou) registration, education, healthcare, and social security systems between rural and urban areas. This unequal distribution of social resources, in turn, prompts a significant migration of rural residents to urban areas in search of better employment opportunities and higher wages. Initially, this mobility can alleviate relative poverty in rural areas due to the differences in income levels and purchasing power, where migrant workers earn higher wages in cities, channeling back into rural regions for consumption. However, as the migration of rural workers to cities continues to evolve and intensify, the portion of urban residents living below the poverty threshold increases. Simultaneously, relative poverty in rural areas is also exacerbated, fundamentally relating to the migration of skilled workers toward cities and the consequent decrease in average income in rural areas. This dual effect can be attributed to the imbalanced economic opportunities and the concentration of wealth in urban areas, leaving both migrant workers and rural areas at an economic disadvantage.

Moreover, in China's major cities, especially in provincial capitals, the prerequisites for household (hukou) registration are considerably strict, demanding specific educational qualifications and income standards. This factor contributes to the phenomenon where cities economically admit rural migrant workers but socially exclude them, impeding the coordinated development of urban and rural areas. In the long run, this situation is unfavorable for managing relative poverty in both urban and rural areas.

### 4.2.3 Educational Factor

Education serves as a fundamental pathway out of poverty. This is underscored by the success of infrastructure development efforts in rural regions of China, which have led to a remarkable increase in the secondary GER, reaching 102.5% by 2020 [24]. However, it is worth noting that the tertiary GER stood merely at 58% in 2020 [25]. In the context of contemporary modern agriculture, skill deficiency is a major barrier preventing impoverished individuals from escaping relative poverty. Those groups with lower levels of knowledge and skills are more likely to encounter challenges of sustaining their livelihoods and are at a higher risk of enduring relative poverty.

Therefore, allocating greater financial resources toward tertiary education becomes crucial to ensuring educational egalitarianism. In China, the stringent requirement of the college entrance examination (gaokao) policy allows only those who reach a certain score threshold to enter tertiary education institutions. Economic growth, however, has not fully addressed the uneven distribution of educational resources between urban and rural areas. The lack of adequately qualified teachers in rural regions challenges rural students' access to education on par with their urban counterparts, thereby intensifying the obstacles they encounter in pursuing higher education.

### 4.2.4 Personal Factor

Due to long-term underprivileged living conditions, the impoverished tend to adopt a lifestyle and values that bear the "stamp of poverty." This is easily passed on to the next generation, hindering the ability of the impoverished to integrate into a rapidly changing society. As a result, impoverished regions experience slower development, and the extremely impoverished population finds it difficult to completely escape poverty, which can eventually lead to a state of mental poverty.

Mental poverty is a significant factor corroding the mindset of the impoverished. Those afflicted with mental poverty often rely on external assistance for a prolonged period, refusing to make contributions through individual labor. Along with China's economic growth, the Chinese government has more funds to subsidize impoverished individuals, leading to their development of mental poverty. Over time, the impoverished population may live solely on government support, lose their existing production skills and social resources, miss new opportunities, find it difficult to adapt to the evolution of the times, and gradually become marginalized groups in societal development.

## 5 Conclusion

The main factor for eliminating absolute poverty in China has been its robust economic growth, which is typically divided into urban and rural economic expansion. Urban economic growth has been primarily driven by the vigorous development of major economic centers, which has increased government tax revenues and funds designated for poverty alleviation. This has, in turn, facilitated infrastructure development and job creation in rural areas. Therefore, impoverished rural residents can either secure better-paying local employment or migrate to urban areas, both effectively elevating their income levels above the poverty threshold. Concurrently, infrastructure development in rural regions contributed to educational enrichment, further enhancing rural residents's eduction return by higher educational attainment.

Although economic growth plays a vital role in alleviating absolute poverty, yet it's important to acknowledge its side effect in increasing income disparity. Such inequality prevents disadvantaged groups from accessing the benefits of economic growth, thereby undermining the poverty-reducing effect of economic growth, specifically in the context of relative poverty alleviation. Furthermore, the exacerbation of relative poverty, in turn, hinders economic development, leading to a vicious cycle of poverty and stunted economic growth. Meanwhile, as a substantial number of rural workers migrated to urban centers, relative poverty in both urban and rural regions escalated, paradoxically as a consequence of the nation's overall economic expansion. Moreover, it is hard for rural residents to escape from relative poverty through higher education, due to the insufficient allocation of educational resources in rural areas. Lastly, the mindset of extremely impoverished individuals discourages them from making individual efforts to extricate themselves from relative poverty and intensifies the challenges that China confronts in its endeavor to reduce relative poverty.

Taking a multidimensional view, poverty encompasses various factors, such as limited access to basic social services, inadequate nutrition, lack of clean water sources, and insufficient healthcare and education. These deficiencies often result in a subjective sense of deprivation and negatively impact overall well-being. In the context of answering the research question, it is apparent that China's economic development from 2011 to 2022 has successfully eliminated absolute poverty and reduced relative poverty. However, this success paradoxically complicates further relative poverty reduction, and could potentially engender a scenario where economic growth is hindered by a surge in relative poverty.

Therefore, strategies to alleviate relative poverty should take regional considerations into account. In middle- and high-income regions, the emphasis should be placed on enhancing residents' capacities and opportunities for higher earnings. Conversely, for impoverished individuals in low-income regions, the focus should be on improving regional economic conditions and fostering an environment conducive to regional economic development, thereby paving a path for their emergence from poverty.